\documentclass[prd,twocolumn,aps,showpacs]{revtex4}
\usepackage[brazil]{babel}
\usepackage[latin1]{inputenc}
\usepackage[T1]{fontenc}
\begin{document}

\title{Comment on "A Lagrangian for DSR Particle and the Role
of Noncommutativity" }

\author{B. F. Rizzuti\footnote{\sf E-mail: brunorizzuti@fisica.ufjf.br On leave
of absence from Instituto de Saúde e Biotecnologia, Universidade
Federal do Amazonas, Coari, Brazil.}} \affiliation{Dept. de
Física, ICE, \\ Universidade Federal de Juiz de Fora, \\
Juiz de Fora, MG, Brazil
\bigskip}


\begin{abstract}

In this work, we comment on two special points in the paper by S.
Ghosh [Phys. Rev. D \textbf{74}, 084019 (2006)]. First of all, the
Lagrangian presented by the author does not describe the
Magueijo-Smolin model of Doubly Special Relativity since it is
equivalent to the Lagrangian of the standard free relativistic
particle. We also show that the introduction of noncommutative
structures is not relevant to the problem of Lorentz covariance in
Ghosh formalism.

\end{abstract}

\pacs{98.80.Cq; 03.30.+p}

\maketitle

In the last few years, there have been a number of papers about a
generalization of Special Relativity (SR). The so-called Doubly
Special Relativity (DSR) \cite{gi,ms,jmls,aad} introduces a new
observer-independent scale, besides $c$, with similar properties
to the latter. There are a number of attractive motivations for
such a modification, listed in \cite{aad}.

In order to introduce the new scale, the construction of DSR
generalizes the conventional energy-momentum dispersion relation
to $p^2=m^2+F(k,E)$, where $m$ is the rest mass; $F$ depends on
the energy and on the new scale $k$, which may be related to the
Planck mass. It is assumed that at the limit $k\rightarrow
+\infty$, $F(k,E)\rightarrow 0$ and we recover the standard
relation $p^2=m^2$.

The most popular model, known as Magueijo-Smolin (MS) DSR
\cite{ms,jmls}, has the dispersion relation
\begin{eqnarray}\label{1}
p^2=m^2(1-\frac{p_0}{k})^2.
\end{eqnarray}
The MS construction is made on conserved energy-momentum space
$\{p_{\mu}\}$, which gives no space-time interpretation for such a
model \cite{mig, agg}.

In order to obtain a space-time picture of the model, a paper was
written \cite{sg}, where a Lagrangian is presented and the author
claims that it corresponds to the MS particle. This comment will
discuss two points of \cite{sg}: first, we will show that the
Lagrangian presented does not correspond to the model of MS DSR.
The other point is that the introduction of noncommutative (NC)
structures is not directly connected to the problem of keeping the
MS dispersion law invariant.

Before starting our comments, it is relevant to say that the work
\cite{sg} was questioned by García in \cite{ag}. This work
contains two particular misleading points. The first one is that
García claimed that the Ghosh Lagrangian can be obtained from the
free relativistic particle Lagrangian by means of an INVERTIBLE
canonical transformation. Nevertheless he uses a SINGULAR
redefinition of momenta,
\begin{equation}\label{a1}
f_{\mu}=\frac{p_{\mu}}{1-\ell p_0}.
\end{equation}
In this case, then, we cannot say that García achieved equivalence
of the free relativistic particle Lagrangian and the Ghosh
Lagrangian that should describe the MS DSR. In particular, the
same redefinition of momenta was already considered in \cite{aad}
to relate Special Relativity to MS DSR.

García also claims that the NC structures introduced by Ghosh are
not correct. The problem is that García's argument is based on a
wrong statement: he says that a canonical transformation preserves
the symplectic structure of the original phase space and does not
preserve the Dirac bracket. But a canonical transformation does
preserve the Dirac bracket \cite{gt}.

Let us now discuss our "first point".

It is important to remember that according to classical mechanics
\cite{hg}, an invertible change of variables in the Lagrangian
action leads to an equivalent description of a given physical
system.

Consider the Lagrangian
\begin{eqnarray}\label{2}
L=m(\dot X^{\mu} \dot X_{\mu})^{\frac{1}{2}}
\end{eqnarray}
which describes the free relativistic particle. We rescale the
coordinates {\it by numbers} as follows:
\begin{eqnarray}\label{3}
X^{\mu} \rightarrow x^{\mu}=(\frac{k^2-m^2}{k^2}X^0,
\frac{\sqrt{k^2-m^2}}{k}X^i).
\end{eqnarray}
Then the Lagrangian reads
\begin{eqnarray}\label{4}
L=\frac{mk}{\sqrt{k^2-m^2}}(\dot x^{\mu} \dot x_{\mu}+
\frac{m^2}{k^2-m^2}(\eta ^{\mu} \dot x_{\mu})^2)^{\frac{1}{2}},
\end{eqnarray}
where $\eta^{\mu}$ is a fixed four-vector defined by $\eta^0=1$,
$\eta^i=0$. Remember that two Lagrangians differing by a total
derivative term are equivalent. Adding this total derivative term,
\begin{eqnarray}\label{5}
\frac{m^2 k}{k^2-m^2}g_{\mu \nu}\dot x^{\mu} \eta ^{\nu}, \qquad g_{\mu\nu}=(1, -1, -1, -1),
\end{eqnarray}
to $L$, we arrive at the Lagrangian proposed in~\cite{sg},
\begin{eqnarray}\label{6}
\label{6} L_{SG}&=&\frac{mk}{\sqrt{k^2-m^2}}[g_{\mu \nu}\dot
x^{\mu} \dot x^{\nu}+ \frac{m^2}{k^2-m^2}(g_{\mu \nu}\dot x^{\mu}
\eta^{\nu})^2 ]^{\frac{1}{2}} \nonumber
\\
&-&\frac{m^2k}{k^2-m^2}g_{\mu \nu}\dot x^{\mu} \eta^{\nu}.
\end{eqnarray}
In summary, the Ghosh Lagrangian is the relativistic particle
Lagrangian (\ref{2}) written in terms of rescaled coordinates
(\ref{3}). In contrast, the MS model (as well as other DSR models
\cite{aad, aad1, aad2}) is related to the standard relativistic
particle formulation by {\it singular} transformation.

"Second point".

Here we first point out that, due to the equivalence with the
relativistic particle Lagrangian (\ref{2}), there is no special
problem with relativistic invariance of the action (\ref{6}).

Ghosh notes that the Lorentz generators in the standard
realization
\begin{eqnarray}\label{6_1}
J_{\mu \nu}=x_{\mu}p_{\nu}-x_{\nu}p_{\mu},
\end{eqnarray}
have non-vanishing Poisson brackets with the MS dispersion
relation
\begin{eqnarray}\label{7}
\{J_{\mu \nu},p^2-m^2(1-\frac{(\eta p)}{k})^2\} = \nonumber
\\
-\frac{2}{k}(1-\frac{(\eta
p)}{k})(\eta_{\mu}p_{\nu}-\eta_{\nu}p_{\mu}),
\end{eqnarray}
and concludes that there is a problem with the Lorentz covariance
of his formalism. His remedy is to add the gauge condition
$(xp)=0$  to the MS first-class constraint (\ref{1}) and to
construct the corresponding Dirac bracket. Since the Dirac bracket
of a constraint with any quantity vanishes, one has, in particular
(we work with $J_{\mu \nu}$ since the new generators $\tilde
J_{\mu \nu}$ introduced by Ghosh, identically coincide with
$J_{\mu \nu}$),
\begin{eqnarray}\label{10}
\{J_{\mu \nu},p^2-m^2(1-\frac{(\eta p)}{k})^2\}_{DB}\approx 0.
\end{eqnarray}

The problem here is that neither the Lagrangian action, nor the
corresponding Hamiltonian action are guaranteed to be invariant
under the transformations defined with the help of the Dirac
bracket instead of the Poisson one.

We point out once again that this last bracket is zero just as a
consequence of the Dirac procedure. So, it is not necessary to
introduce NC variables to arrive at the vanishing Dirac bracket of
the generator with the MS constraint.

Hence we have shown that the Ghosh Lagrangian is just the free
relativistic particle Lagrangian, in a different parametrization.
We note that, in contrast to \cite{ag}, we connected the Ghosh
Lagrangian with the standard free particle Lagrangian by an
invertible transformation. We also have shown that the NC
structures introduced in \cite{sg} are not connected to the
problem of keeping MS dispersion relation invariant. We point out
that the problem of consistent construction of DSR models in
configuration space is still an open issue. We observe that new
ideas have appeared in order to solve this problem
\cite{smig,fsl}.

The author would like to thank FAPEMIG for their financial support
and Prof. A. A. Deriglazov for helpful discussions.

\end{document}